# Na Site Doping a Pathway for Enhanced Thermoelectric Performance in $Na_{1-x}CoO_2$; The Case of Gd and Yb Dopants

M. Hussein N. Assadi[*]

*School of Materials Science and Engineering, UNSW Australia, Sydney, NSW, 2052, Australia*

Doping is considered to be the main method for improving the thermoelectric performance of layered sodium cobaltate ($Na_{1-x}CoO_2$). However, in the vast majority of past reports, the equilibrium location of the dopant in the $Na_{1-x}CoO_2$'s complex layered lattice has not been confidently identified. Consequently, a universal strategy for choosing a suitable dopant for enhancing $Na_{1-x}CoO_2$'s figure of merit is yet to be established. Here, by examining the formation energy of Gd and Yb dopants in $Na_{0.75}CoO_2$ and $Na_{0.50}CoO_2$, we demonstrate that in an oxygen poor environment, Gd and Yb dopants reside in the Na layer while in an oxygen rich environment these dopants replace a Co in $CoO_2$ layer. When at Na layer, Gd and Yb dopants reduce the carrier concentration via electron-hole recombination, simultaneously increasing the Seebeck coefficient ($S$) and reducing electric conductivity ($\sigma$). Na site doping, however, improves the thermoelectric power factor (PF) only in $Na_{0.50}CoO_2$. When replacing a Co, these dopants reduce $S$ and PF. The results demonstrate how thermoelectric performance critically depends on the synthesis environment that must be fine-tuned for achieving any thermoelectric enhancement.

## INTRODUCTION

Sodium cobaltate ($Na_{1-x}CoO_2$) is an interesting compound that exhibits rich magnetic and structural phase diagrams [1-3] and possesses a relatively high thermoelectric figure of merit ($ZT$) [4] at higher temperatures. As shown in Figure 1(a), the $Na_{1-x}CoO_2$ lattice is made of stacked alternating Na layers and edge-sharing $CoO_2$ octahedra. The mixture of $Co^{3+}$ and $Co^{4+}$ ions in the $CoO_2$ layer in sodium deficient $Na_{1-x}CoO_2$ generates a Seebeck potential through spin entropy flow [5, 6]. The Seebeck coefficient ($S$) of polycrystalline $Na_{0.75}CoO_2$ at 800 K is within the range of ~132 $\mu V K^{-1}$ and ~143 $\mu V K^{-1}$ depending on the synthesis method [7-9]. This $S$ value is comparable with that of other thermoelectric oxides; For instance, ZnO:Al and $CaMoO_3$:Gd have $S$ values of ~ −110 $\mu V K^{-1}$ [10] and ~ −225 $\mu V K^{-1}$ [11] respectively at the same temperature (For comprehensive reviews see Refs. [12] and [13]). In $Na_xCoO_2$, phonons, however, are diffusively scattered by the $Na^+$ ions which are mobile at room temperature [14, 15] reducing the lattice thermal conductivity ($\kappa_L$) [16-20] to ~ 0.01 $W\,cm^{-1}K^{-1}$ at ~1000 K [21]. This $\kappa_L$ value is considerably smaller than that of most oxides in which the dominance of covalent bonding causes relatively high $\kappa_L$; for instance; ZnO has a $\kappa_L$ value of ~1.25 $W\,cm^{-1}K^{-1}$ at ~1000 K [22].

Given the phenomenal and promising low $\kappa_L$ in $Na_{1-x}CoO_2$, doping has extensively been used to improve the Seebeck coefficient of $Na_{1-x}CoO_2$ further with the ambition of bringing its $ZT$ to values above unity [7, 8, 21, 23-32]. Nonetheless, dopants have usually been chosen based on the mere nominal oxidation state, atomic mass considerations and the solubility limits of the applied synthesis method [33]. An overlooked issue, nonetheless, is that cationic dopants can be principally substituted for either Na or Co ions. Identifying the exact location of the cationic dopant in $Na_{1-x}CoO_2$'s lattice under a specific condition requires characterisations sensitive to the local chemical environments such as X-ray absorption spectroscopy and neutron diffraction which are most often absent from the existing reports so far. Consequently, despite the large volume of research on doped $Na_{1-x}CoO_2$, the experimental advancement in the doped $Na_{1-x}CoO_2$ has been mainly guided by the approximate guesswork rather comprehensive and strategic insight of how dopants influence the $ZT$ through structure-property relationship. As a result, not only the initial ambition of achieving a $ZT$ comfortably greater than one has not been realised yet but also many theoretically interesting questions have remained unanswered.

In this work, therefore, we examine the energetics and the electronic structure of Gd and Yb doped $Na_{1-x}CoO_2$ for $x = 0.25$ and 0.5. Based on the formation energy calculations, we demonstrate that the location of Yb and Gd dopants in $Na_{1-x}CoO_2$ critically depends on the synthesis environment, i.e. these dopants reside in different lattice sites depending on the O partial pressure during the synthesis. The insight obtained here complements our previous work that demonstrated that the location of dopants such as Cu and Au also depends on the Na content of $Na_{1-x}CoO_2$ [34]. One of the main conclusions that we draw in this work is that the location of cationic dopants in $Na_{1-x}CoO_2$ needs to be investigated critically and the simplistic assumption based on matching ionic radii in determining dopant location can be misleading at times.

## COMPUTATIONAL SETTINGS

Total energy density functional calculations were performed using the plane-wave and on-the-fly generated ultra-soft pseudopotential [35] approach, as implemented in CASTEP [36-38]. Ceperley and Alder's local density approximation was used for the exchange-correlation term in the Hamiltonian [39]. Energy cut-off was set to 517 eV, and the $k$-point mesh was set to generate a k point separation of 0.05 Å$^{-1}$ for oxides and 0.01 Å$^{-1}$ for metals. The density-mixing scheme was applied for electronic minimisation during which the spin of all atoms was initiated based on formal values and then allowed to relax.

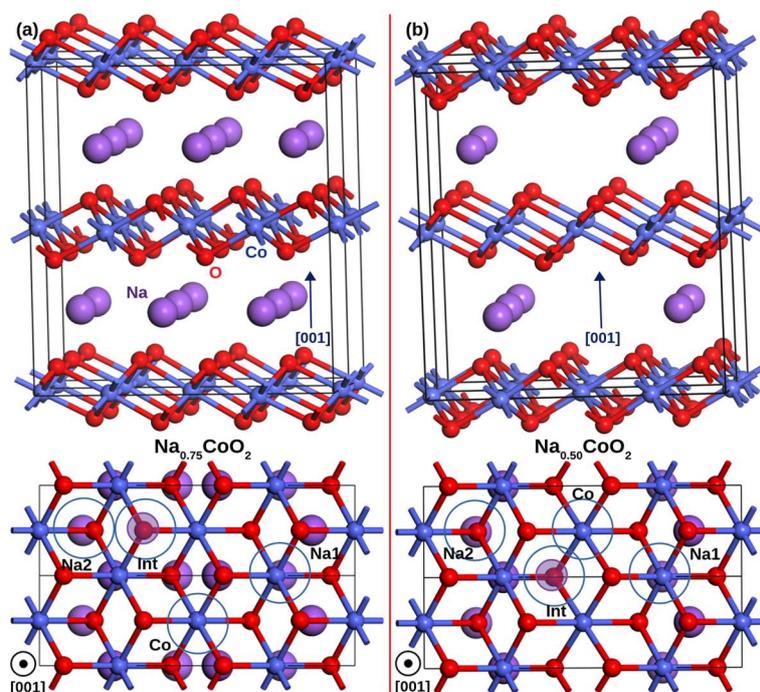

Figure 1. A side (upper panels) and top (lower panels) preview of the $Na_{0.75}CoO_2$ and $Na_{0.50}CoO_2$ supercells is provided in (a) and (b), respectively. Co and O ions occupy the Wyckoff $2a$ and $4f$ sites of the hexagonal lattice structure respectively. In $Na_{0.75}CoO_2$ compound, one-third of the Na ions occupy $2b$ (Na1) sites which share basal coordinates with Co and two-thirds of Na ions occupy $2d$ (Na2) sites which share basal coordinates with O. In $Na_{0.50}CoO_2$ half of the Na ions occupy Na1 site while the other half occupy Na2 sites. The halftone circles represent the location of the interstitial dopants that is vacant in the undoped compounds.

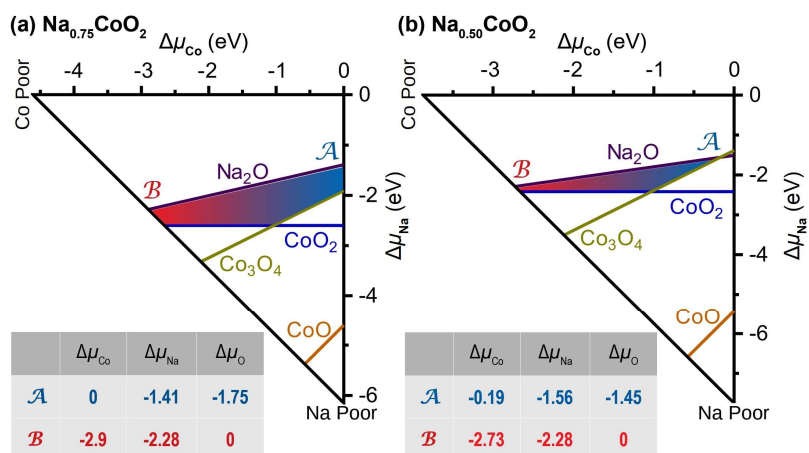

Figure 2. The accessible ($\Delta\mu$) chemical potential range. The triangle vertices are determined by the formation enthalpy of (a) $Na_{0.75}CoO_2$ and (b) $Na_{0.50}CoO_2$, respectively. Limits are imposed by the formation of competing binary phases and result in the shaded stable region. $\Delta\mu$ values are given for points $A$ and $B$ in eV.

The scalar relativistic treatment based on Kœlling-Harmon approximation of Dirac's equation was also applied [40]. LDA+$U$ correction based on a simplified and rotationally invariant approach was applied to Co 3d and Gd/Yb 4f electrons [41]. The default $U$ values of 2.5 eV for Co and 6.0 eV for rare earth dopants were selected for which a full justification is provided in Figure S1 of the Supplementary Information. The lattice parameters of a fully optimised primitive unit cell of $Na_1CoO_2$ was found to be 2.87 Å for $a$ and 10.90 Å for $c$ reasonably matching the experimental values [42]. The difference was only 0.07% for $a$ and −1.49% for $c$. Then a $2a\times4a\times1c$ $Na_1CoO_2$ supercell was constructed for studying the doped compounds. Four and then eight out of the sixteen original Na ions were removed according to the previously established patterns [43], shown in Figure 1, to create a $Na_{12}Co_{16}O_{32}$ and $Na_8Co_{16}O_{32}$ supercells for $Na_{0.75}CoO_2$ and $Na_{0.50}CoO_2$ compounds, respectively. Further details regarding the convergence with respect to the supercell size is provided in Table S1 and Figure S2 of the Supplementary Information. The oxidation state of the dopants and the Co ions was estimated from the magnetisation calculated by Mulliken population analysis and examining the partial density of states. One, however, should note that due to partial covalency in Co—O bond, Co ion magnetisation is slightly smaller than what is anticipated from Hund's rule [44]. The accuracy of Mulliken population analysis was cross-examined with Hirshfeld charge analysis for witch the results are provided in Table S2. It was found that Mulliken population analysis provides a robust description of charge localisation in doped $Na_{1-x}CoO_2$ compounds.

**RESULTS AND DISCUSSION**

Dopants' formation energy ($E^f$) was calculated for four possible replacement configurations. In the first configuration, the dopant M replaced a Co ion creating an $M_{Co}$ configuration. In the second instant, M occupied an interstitial site in the Na layer, creating a $M_{Int}$ configuration. Third, dopant M replaced a Na ion at Na1 site, creating an $M_{Na1}$ configuration. Finally, M replaced a Na ion at Na2 site, creating an $M_{Na2}$ dopant. As demonstrated in Figure 1, Na1 shares the basal coordinates with Co while Na2 shares the basal coordinates with O. The formation energy ($E^f$) was calculated using the standard procedure as described by the following equation [45]:

$$E^f = E^t(Na_{1-x}CoO_2:M) + \mu_\alpha - E^t(Na_{1-x}CoO_2) - \mu_M \quad (1).$$

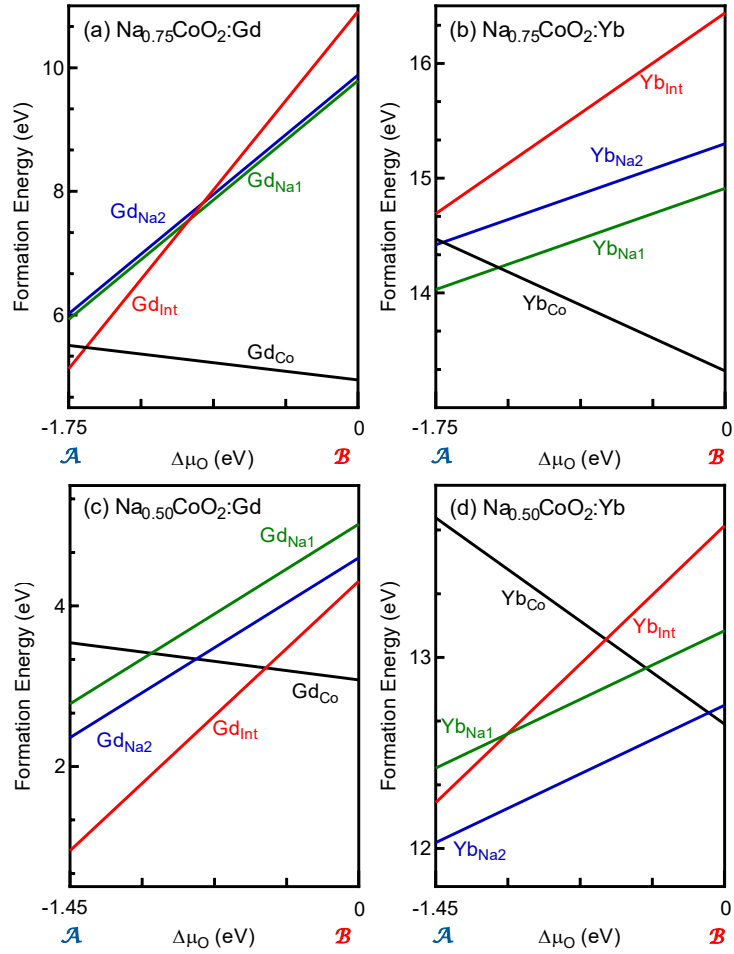

Figure 3. The formation energy of Yb and Gd dopants in $Na_{0.75}CoO_2$ and $Na_{0.50}CoO_2$ compounds, respectively. Notions $A$ and $B$ correspond to the points that are marked similarly in Figure 2, corresponding to O poor an O rich environments respectively.

Here, $E^t(Na_{1-x}CoO_2:M)$ is the total energy of the $Na_{1-x}CoO_2$ supercell containing the dopant M, and $E^t(Na_{1-x}CoO_2)$ is the total energy of the undoped $Na_{1-x}CoO_2$ supercell. $\mu_\alpha$ and $\mu_M$ are the chemical potentials of the removed and added elements, respectively. The chemical potentials depend on the synthesis environment [46]. To investigate the thermodynamics of dopant solubility in $Na_{1-x}CoO_2$, we first determined the accessible chemical potentials for Na, Co, O and the dopants Gd and Yb. By varying the chemical potentials ($\mu$) by a permissible value of $\Delta\mu$ ($\mu = \mu^0 + \Delta\mu$), we can simulate the effect of varying the oxygen partial pressures and the abundance of constituting elements on the dopants' formation energy and their location in the host lattice. We, therefore, can determine the optimum conditions for Gd and Yb doping that may enhance the thermoelectric performance. The first constraint on the chemical potentials is set by the enthalpy of the $Na_{1-x}CoO_2$:

$$(1-x)\Delta\mu_{Na} + \Delta\mu_{Co} + 2\Delta\mu_O = \Delta H^f(Na_{1-x}CoO_2) \quad (2)$$

in which $\Delta H^f(Na_{1-x}CoO_2)$ is the DFT formation enthalpy of $Na_{1-x}CoO_2$. Furthermore, to avoid precipitation into solid elemental Co, Na, and the release of gaseous $O_2$, we also require:

$$\Delta\mu_{Na}, \Delta\mu_{Co}, \Delta\mu_O \leq 0. \quad (3)$$

The chemical potentials are further constrained by the decomposition of $Na_{1-x}CoO_2$ into competing binary compounds such as $Na_2O$, $CoO_2$ and $Co_3O_4$:

$$2\Delta\mu_{Na} + \Delta\mu_O = \Delta H^f(Na_2O), \quad \Delta\mu_{Co} + 2\Delta\mu_O = \Delta H^f(CoO_2), \quad 3\Delta\mu_{Co} + 4\Delta\mu_O = \Delta H^f(Co_3O_4). \quad (4)$$

For Gd and Yb, $\Delta\mu$ was calculated based on:

$$2\Delta\mu_{Gd} + 3\Delta\mu_O = \Delta H^f(Gd_2O_3), \quad (5)$$
$$\Delta\mu_{Yb} + 2\Delta\mu_O = \Delta H^f(YbO). \quad (6)$$

Here $Ia\bar{3}$ $Gd_2O_3$ and $Fm\bar{3}m$ YbO are the most stable Yb and Gd oxides. Finally, the chemical potentials were set equal to the elemental energy of a given metal ($\mu^0$) plus the corresponding $\Delta\mu$.

As shown in Figure 2, we found that the major limiting phases are $Na_2O$ in Na rich environment and $CoO_2$ and $Co_3O_4$ in Co rich environment for both $Na_{0.75}CoO_2$ and $Na_{0.5}CoO_2$. CoO was not a limiting phase. Furthermore, the available range of the chemical potential was relatively limited by the permissible range of $\Delta\mu_{Na}$, resulting in a narrow strip that had a wider range for $\Delta\mu_{Co}$ and $\Delta\mu_O$. Due to the narrow strip of permissible $\Delta\mu$ values, we only present two extremes when discussing the formation energy. These extremes are marked with $A$ for O poor environment and $B$ for O rich environment in Figure 2 (justification is provided in Figure S3 and Table S3 of the Supplementary Information).

Figure 3 presents the formation energy of the dopants. Under O poor environment, In $Na_{0.50}CoO_2$ and $Na_{0.75}CoO_2$, both Gd and Yb dopants reside in the Na layer. In the case of Gd doped $Na_{0.75}CoO_2$, $Gd_{Int}$ with an $E^f$ of 5.13 eV was the most stable configuration followed by $Gd_{Co}$ with an $E^f$ of 5.51 eV. In the case of Yb doped $Na_{0.75}CoO_2$, $Yb_{Na1}$ with an $E^f$ of 14.03 eV was the most stable configuration followed by $Yb_{Na2}$ with an $E^f$ of 14.42 eV. In the case of Gd doped $Na_{0.50}CoO_2$, the most stable configuration was $Gd_{Int}$ with an $E^f$ of 0.95 eV followed by $Gd_{Na2}$ with an $E^f$ of 2.36 eV while in the case of Yb doped $Na_{0.50}CoO_2$, the most stable configuration was $Yb_{Na2}$ with an $E^f$ of 12.03 eV followed by $Yb_{Int}$ with an $E^f$ of 12.24 eV.

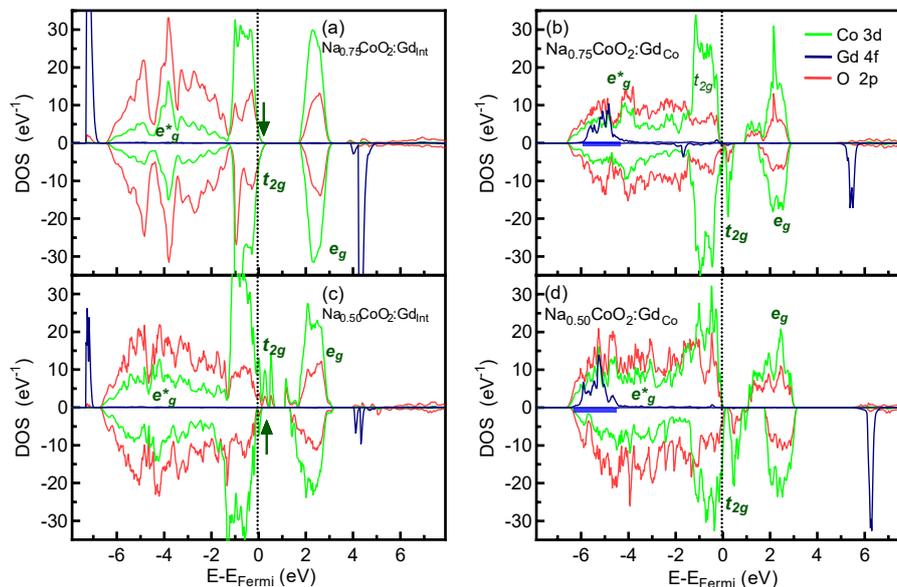

Figure 4. Partial density of states of the most stable doping configurations in O poor environment (left column) and O rich environment (right column) for Gd doped $Na_{0.75}CoO_2$ (top panels) and $Na_{0.50}CoO_2$ (bottom panels). Blue, green and red lines represent Gd 4f, Co 3d and O 2p states, respectively.

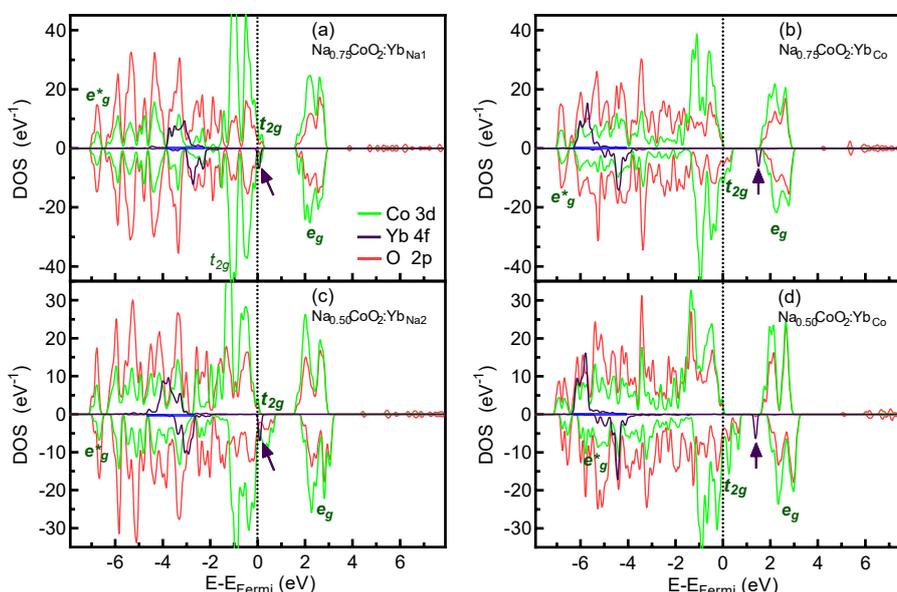

Figure 5. Partial density of states of the most stable doping configurations in O poor environment (left column) and O rich environment (right column) for Yb doped $Na_{0.75}CoO_2$ (top panels) and $Na_{0.50}CoO_2$ (bottom panels). Purple, green and red lines represent Yb 4f, Co 3d and O 2p states, respectively.

Under O rich environment, the sequence of the stabilisation was, however, different. In both $Na_{0.50}CoO_2$ and $Na_{0.75}CoO_2$ compounds; Gd and Yb dopants replace a Co ion in the host lattice. For Gd doped $Na_{0.75}CoO_2$, $Gd_{Co}$ with an $E^f$ of 4.95 eV had the lowest energy followed by $Gd_{Na1}$ with an $E^f$ of 9.79 eV. For Yb doped $Na_{0.75}CoO_2$, $Yb_{Co}$ with an $E^f$ of 13.32 eV had the lowest energy followed by $Yb_{Na1}$ with an $E^f$ of 14.91 eV. In the case of Gd doped $Na_{0.50}CoO_2$, $Gd_{Co}$ was the most stable configuration with an $E^f$ of 3.08 eV followed by $Gd_{Int}$ with an $E^f$ of 4.31 eV while in the case of Yb doped $Na_{0.50}CoO_2$, the most stable $Yb_{Co}$ configuration had an $E^f$ of 12.69 eV followed by $Yb_{Na2}$ with an $E^f$ of 12.75 eV.

Figure 4 presents the partial density of states (PDOS) of the stable Gd doped configurations in O poor environment, i.e. $Na_{0.75}CoO_2:Gd_{Int}$ and $Na_{0.50}CoO_2:Gd_{Int}$, and O rich environment, i.e. $Na_{0.75}CoO_2:Gd_{Co}$ and $Na_{0.50}CoO_2:Gd_{Co}$. In all of these configurations, Gd's spin-up channel is completely filled while the spin-down channel is empty indicating that Gd adapts an oxidation state of +3 independent from O partial pressure or Na content. Mulliken population analysis also shows a ~7 $\hbar/2$ magnetisation for Gd in all of these configurations indicating a [Xe] $4f^75d^06s^0$ electronic configuration, that is $Gd^{3+}$. For $Gd_{Int}$ which is stable at O poor environment, the three electrons introduced by the interstitial $Gd^{3+}$ reduce three of the $Co^{4+}$ ions in the $Na_{0.75}CoO_2:Gd_{Int}$ and $Na_{0.50}CoO_2:Gd_{Int}$ to $Co^{3+}$. As a result, $Na_{0.75}CoO_2:Gd_{Int}$ has only one spin bearing Co with a magnetisation of 0.76 $\hbar/2$ (undoped $Na_{0.75}CoO_2$ has a total spin of 3.96 $\hbar/2$ borne on four out of the 16 Co ions in the supercell commensurate with four low-spin $Co^{4+}$ in tetrahedral coordination i.e.

$t_{2g}^5 e_g^0$), while $Na_{0.50}CoO_2$:$Gd_{Int}$ has five $Co^{4+}$ ions with a total magnetisation of 4.37 $\hbar/2$. As marked with arrows in Figure 4(a) and (c), the introduction of $Gd_{Int}$ has, indeed, reduced the peak height of empty $t_{2g}$ states that indicating a reduction of $Co^{4+}$ concentration. In the case of $Na_{0.75}CoO_2$:$Gd_{Co}$, the stable configuration at O rich environment, the total spin borne on Co ions was 3.80 $\hbar/2$, very close to the value of the undoped compound, indicating that $Gd^{3+}$ has replaced a $Co^{3+}$ ion leaving the four $Co^{4+}$ ions unaltered. The same argument holds for $Na_{0.5}CoO_2$:$Gd_{Co}$ as the total spin borne on Co ions is 7.2 $\hbar/2$ indicating eight $Co^{4+}$ in the supercell.

The PDOS of Gd doped compounds has some other noticeable features. For instance, as marked with blue bars in Figure 4(b) and (d), the filled spin-up Gd 4f states of $Gd_{Co}$ spread over the range of $-6$ eV$<E_{Fermi}<-4$ eV, while the same states are sharply localised at $\sim-7$ eV for $Gd_{Int}$ for both Gd doped $Na_{0.75}CoO_2$ and $Na_{0.50}CoO_2$ compounds. The spread of the $Gd_{Co}$ 4f states, although not considered full delocalisation, demonstrate the effect of O coordination and its hybridisation with 4f states. Furthermore, the Gd 4f and $Co^{4+}$ 3d ions have parallel spins in all considered compounds in Figure 4 except for $Na_{0.50}CoO_2$:$Gd_{Int}$ for which Gd 4f and $Co^{4+}$ 3d states are of antiparallel spins.

Figure 5 presents the PDOS of Yb doped compounds that are the most stable at O rich and O poor environments, respectively. For compounds that are most stable at O poor environment, that is $Yb_{Na1}$ in $Na_{0.75}CoO_2$ and $Yb_{Na2}$ in $Na_{0.5}CoO_2$, Yb 4f states have a narrow empty peak located just above the Fermi level, marked with arrows in Figure 5(a) and (c), indicating that Yb dopant had an oxidation state of $+3$ i.e. [Xe] $4f^{13}6s^0$. Considering the remaining spin borne on Co ions which was 1.96 $\hbar/2$ in $Na_{0.75}CoO_2$:$Yb_{Na1}$ and 5.12 $\hbar/2$ in $Na_{0.50}CoO_2$:$Yb_{Na2}$, we conclude that $Yb^{3+}$ dopants replacing $Na^{1+}$ introduce two electrons into the compounds reducing two $Co^{4+}$. Moreover, Co substituting Yb dopants that are stable in O rich environment also adapted $+3$ oxidation state as demonstrated by the empty 4f states marked with arrows in Figure 5(b) and (d) and the spin borne on Yb ions of 0.95 $\hbar/2$. In these latter cases, the spin borne on Co ions did not differ much from the values borne in undoped compounds, indicating a $Yb^{3+}$ substituting for a $Co^{3+}$.

We just demonstrated how the location of both Gd and Yb dopants which critically depend on the O partial pressure, influence the electronic structure of the host material. Now let's examine the implication of dopant's location on the thermoelectric performance of Gd and Yb doped $Na_{1-x}CoO_2$ in terms of Seebeck coefficient ($S$), carrier concentration ($n$), conductivity ($\sigma$) and power factor (PF = $S^2\sigma$). The high-temperature Seebeck coefficient in $Na_{1-x}CoO_2$ can be explained by Koshibæ's equation [47, 48] for strongly correlated materials which is a modified form of Heikes formula [49, 50]:

$$S(T \to \infty) = -\frac{k_B}{e}\ln\left[\frac{g(Co^{4+})}{g(Co^{3+})}\frac{n_{Co^{4+}}}{n_{Co^{3+}}}\right] \quad (7).$$

Here $k_B$ is the Boltzmann constant and $e$ is the electron charge, $g$ equals to the different possible ways in which electrons can be arranged in the orbitals of $Co^{3+}$ and $Co^{4+}$ ions, and $n$ is the concentration of a given species of Co. $g$ can be expressed as the product of spin degeneracy ($g_s$) and orbital ($g_o$) degeneracy: $g = g_s \cdot g_o$. $g_s$ equals to $2\zeta+1$ where $\zeta$ is the ions' total spin number while $g_o$ is the number of valid permutations for distributing the electrons across its orbitals. Assuming that Co ions take low spin state in $Na_{1-x}CoO_2$ ($\zeta = 0$ for $Co^{3+}$ and $\zeta = \frac{1}{2}$ for $Co^{4+}$), we obtain $g(Co^{4+}) = 6$ and $g(Co^{3+}) = 1$. By substituting these values in the modified Koshibæ's formula for $S$ of an electron hopping from a $Co^{3+}$ ion to a $Co^{4+}$ ion, we obtain values of $S = 249$ $\mu$V K$^{-1}$ for $Na_{0.75}CoO_2$ and $S = 154$ $\mu$V K$^{-1}$ for $Na_{0.50}CoO_2$. It should be noted that Koshibæ's formula was obtained by solving the transport problem for a strongly correlated oxide using a Hubbard model at an infinite temperature [51]. The yielded results are, nonetheless, valid for doped $Na_{1-x}CoO_2$, as these compounds are generally intended for waste heat recovery at temperatures higher than $\sim$700 K [12]. For further details see Figure S4 of the Supplementary Information.

Furthermore, we can approximate the conductivity as a function of $Co^{4+}$ concentration by assuming that the carrier mobility remains 1.0 cm$^2$V$^{-1}$s$^{-1}$ for doped $Na_{1-x}CoO_2$ [52] for which full details are provided in Table S4. This approximation is somehow conservative as dopants in Na layer are generally expected to improve carrier mobility slightly [34]. This approximation gives a conductivity value of $1.03 \times 10^3$ $\Omega^{-1}$cm$^{-1}$ for $Na_{0.75}CoO_2$ and $2.06 \times 10^3$ $\Omega^{-1}$cm$^{-1}$ for $Na_{0.50}CoO_2$. The power factor, which is a parabolic function of $S$ and a linear function of $\sigma$, comes out as 6.392 mW m$^{-1}$K$^{-2}$ for $Na_{0.75}CoO_2$ and 4.912 mW m$^{-1}$K$^{-2}$ in $Na_{0.50}CoO_2$. These values are quite similar to measurements in single-crystal [53, 54] and epitaxial thin film $Na_{1-x}CoO_2$ [55]. The conductivity values, however, are an order of magnitude larger than those measured in polycrystalline samples indicating the significant role of carrier scattering at grain boundaries [56].

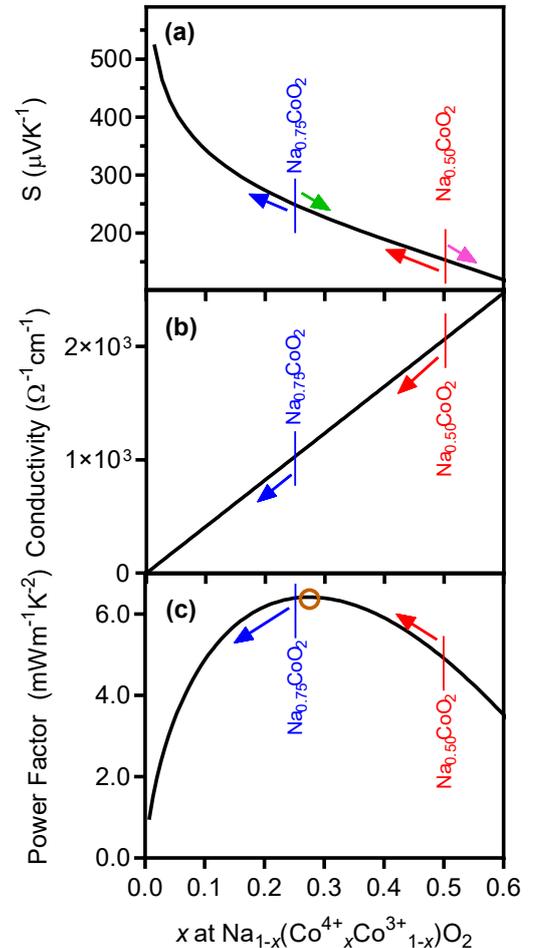

Figure 6. (a) Seebeck coefficient ($S$) based on Koshibæ's equation, (b) conductivity ($\sigma$) and (c) the power factor in $Na_{1-x}CoO_2$ as a function of $Co^{4+}$ concentration. Green and pink arrows indicate the change in $S$ for cobalt site doping. The blue and red arrows indicate the change of $S$, $\sigma$ and PF for Na site doping.

In O poor environment, $Gd_{Int}$ is the most stable configuration in both $Na_{0.75}CoO_2$ and $Na_{0.50}CoO_2$. Interstitial $Gd^{3+}$ introduces three electrons that recombine with three holes of the host compound, reducing the concentration of $Co^{4+}$. As marked with red and blue arrows in Figure 6(a), such a reduction in $Co^{4+}$ concentration increases $S$ to 388 $\mu VK^{-1}$ in $Na_{0.75}CoO_2$:$Gd_{Int}$, and to 222 $\mu VK^{-1}$ in $Na_{0.50}CoO_2$:$Gd_{Int}$. The reduction in $Co^{4+}$ concentration also decreases carrier concentration ($n$) and conductivity ($\sigma$). As marked with red and blue arrows in Figure 6(b), $Gd_{Int}$ reduces $\sigma$ to $2.58 \times 10^2$ $\Omega^{-1}cm^{-1}$ in $Na_{0.75}CoO_2$:$Gd_{Int}$ and to $1.29 \times 10^3$ $\Omega^{-1}cm^{-1}$ in $Na_{0.50}CoO_2$:$Gd_{Int}$. As shown in Figure 6(c), the power factor, decreases to 3.873 $mWm^{-1}K^{-2}$ for $Na_{0.75}CoO_2$:$Gd_{Int}$ while it increases to 6.367 $mWm^{-1}K^{-2}$ in $Na_{0.50}CoO_2$:$Gd_{Int}$. In O poor environment, Yb dopants replace an existing Na. In this case, each Yb dopant introduces two electrons. As a result, the reduction in $Co^{4+}$ concentration and conductivity is less dramatic than the case of Gd doping. For $Na_{0.75}CoO_2$:$Yb_{Na1}$, $S$, $\sigma$ and power factor, therefore, were calculated to be 322 $\mu VK^{-1}$, $5.15 \times 10^2$ $\Omega^{-1}cm^{-1}$ and 5.344 $mWm^{-1}K^{-2}$, respectively. For $Na_{0.50}CoO_2$:$Yb_{Na2}$, on the other hand, $S$, $\sigma$ and power factor, were calculated to be 198 $\mu VK^{-1}$, $1.55 \times 10^3$ $\Omega^{-1}cm^{-1}$ and 6.085 $mWm^{-1}K^{-2}$ respectively.

$Gd^{3+}$ and $Yb^{3+}$ dopants, when substituting for $Co^{3+}$ as in O rich environment, do not change the carrier (hole) concentration. However, they change the dynamics of spin entropy flow. As shown in Figure 4(b) and (d) and Figure 5(b) and (d), the 4f states of Gd and Yb dopants gravitate towards the bottom of the valence band hybridising with O 2p and Co's bonding $e_g^*$ states. The electric conduction and therfore the spin entropy flow is facilitated by the electrons hopping from a full $t_{2g}^6$ states of a $Co^{3+}$ ion to a singly vacant $t_{2g}^5$ states of a $Co^{4+}$ which all are located within ~1 eV of the Fermi level. Consequently, $Gd_{Co}$ and $Yb_{Co}$, in practice, reduce the $Co^{3+}$ sites available for conduction, increasing the overall concentration of $Co^{4+}$. Taking the reduced number of $Co^{3+}$ sites into account, as marked with green and pink arrows in Figure 6(a), according to Koshibæ's equation, $S$ is reduced to 241 $\mu VK^{-1}$ for $Na_{0.75}CoO_2$ doped with $Gd_{Co}$ and $Yb_{Co}$ and to 142 $\mu VK^{-1}$ for $Na_{0.50}CoO_2$ doped with $Gd_{Co}$ and $Yb_{Co}$. The carrier concentration, on the other hand, is not altered by Co side doping as the concentration of hole bearing $Co^{4+}$ does not change by either $Gd_{Co}$ or $Yb_{Co}$ doping. The net result for Co site doping that which is prevalent in O rich environment is, therefore, a net decrease in the power factor.

Earlier experiments have confirmed the possibility of doping in $Na_{0.50}CoO_2$ with Gd and Yb [23, 57]. In the case of Yb doping in $Na_{0.5}CoO_2$, synthesised through solid-state reaction, 5% Yb doping decreased $\sigma$ by ~25% compared to the undoped compound over the temperature range of 400 - 1000 K (from ~$2.94 \times 10^2$ $\Omega^{-1}cm^{-1}$ to $2.27 \times 10^2$ $\Omega^{-1}cm^{-1}$). Yb doping, on the other hand, slightly increased $S$ from ~185 $\mu VK^{-1}$ to ~205 $\mu VK^{-1}$ at 800 K [23]. As a result, Yb doping increased the PF to ~1.5 $mWm^{-1}K^{-2}$ from the ~1.2 $mWm^{-1}K^{-2}$ of the undoped sample. The effect of Yb doping on the thermoelectric $Na_{0.50}CoO_2$ is in general agreement with our theoretical prediction. One, however, should note that, as shown in Figure 6(c), the PF as a function of carrier concentration has a maximum of 6.416 $mWm^{-1}K^{-2}$ at a $Co^{4+}$ concentration of 0.275. The hypothetical compound $Na_{0.725}CoO_2$ would have the maximum attainable power factor. This composition is, however, above the convex hull of the $Na_{1-x}CoO_2$ phase diagram [58, 59]. As a result, altering $Co^{4+}$ concentration through doping in a stable $Na_{1-x}CoO_2$ composition is the only way to attain the maximum PF value. Consequently, electron doping, such as the case of $Yb_{Na}$ and $Gd_{Int}$, in the Na layer in $Na_{0.50}CoO_2$ moves the PF towards that maximum while electron doping in $Na_{0.75}CoO_2$ takes the thermopower away from this maximum.

$Ca_3Co_4O_{9+\delta}$ is another compound with many similarities to $Na_{1-x}CoO_2$ in which the Seebeck effect originates from the spin entropy flow from $Co^{3+}$ to $Co^{4+}$ ions. Similar to our prediction for Gd and Yb doping in $Na_{0.50}CoO_2$, Gd doping at 10% in $Ca_3Co_4O_{9+\delta}$ was found to decrease $\sigma$ from ~100 $\Omega^{-1}cm^{-1}$ to ~67 $\Omega^{-1}cm^{-1}$ and to increase $S$ from 125 $\mu VK^{-1}$ to 145 $\mu VK^{-1}$ at 400 K [60]. Similarly, ~13% Yb doping in $Ca_3Co_4O_{9-\delta}$ was found to decrease $\sigma$ to ~45 $\Omega^{-1}cm^{-1}$ which is half of the value of the undoped compound and to raise $S$ to 155 $\mu VK^{-1}$ at 400 K [61]. Judging from the combined effect on conductivity and Seebeck coefficient, these dopants, most likely, occupy a Ca site in calcium cobaltate.

## CONCLUSIONS

Through density functional calculations with LDA+$U$ formalism, we showed that the formation energy of Gd and Yb dopants in $Na_{0.75}CoO_2$ and $Na_{0.50}CoO_2$ was critically sensitive to the synthesis conditions. In an oxygen poor environment, Gd and Yb dopants preferred to occupy a spot in the Na layer while in oxygen rich environment these dopants replaced a Co. Since Gd and Yb have higher oxidation state than Na, when at Na layer, these dopants reduce the carrier concentration and the electric conductivity via electron-hole recombination and increase the Seebeck coefficient at the same time. The thermoelectric power factor, however, improves only for doped $Na_{0.50}CoO_2$ for which the increased Seebeck coefficient supersedes the reduction in the electric conductivity. When replacing a Co, Gd and Yb dopants reduce the Seebeck coefficient while leaving the electric conductivity unchanged resulting in a net reduction in the power factor.

## ACKNOWLEDGEMENTS

Computational resources were provided by the Integrated Materials Design Centre at UNSW Australia.

Supplementary Information

# Na site doping a pathway for enhanced thermoelectric performance in Na$_{1-x}$CoO$_2$; The case of Gd and Yb dopants

M. Hussein N. Assadi

*School of Materials Science and Engineering, UNSW Australia, Sydney, NSW, 2052, Australia*
h.assadi.2008@ieee.org

## Table of Contents

## The Choice of *U* Values

Here we examined how the application of the LDA+$U$ can alter the conclusions drawn in the Article. Since $U$ can not be usually determined *a priori*, selecting a proper $U$ value for d and f electrons always presents a challenge.[1] $U$ values, therefore, are chosen to replicate some experimental parameter, often the bandgap from optical measurements. This approach is, however, not applicable to Na$_{1-x}$CoO$_2$ as this compound is metallically conductive and thus opaque. As a result, $U$ value for sodium cobaltate is chosen in a way that reproduces Co ions' charge disproportionation.[2,3] In the literature, there are plenty of different values reported for Co that achieve this goal.[4,5] Furthermore, the choice of $U$ also depends on the type of the pseudopotential.[6,7] In our work, we applied the default values implemented in Materials Studio for Co and Gd/Yb dopants; $U$ = 2.5 eV for Co 3d electrons and $U$ = 6 eV for Gd/Yb 4f electrons. Figure S1(a) shows the partial density of states (PDOS) of Na$_{0.75}$CoO$_2$:Yb calculated with LDA functional. Here, we do not see the charge disproportionation as all Co ions had uniform 3d population of 5.75 $e$. Figure S1(b) shows the PDOS of the same system calculated with $U$ = 2.5 eV for Co 3d electrons and $U$ = 6 eV for Yb 4f electrons. Here, we see that the LDA+$U$ method results in two distinct types of Co ions: the Co$^{3+}$ ($t_{2g}^6 e_g^0$) and the Co$^{4+}$ ($t_{2g}^5 e_g^0$). Furthermore, now, the energy gap between Co$^{3+}$ filled $t_{2g}^6$ and empty $e_g^0$ is ~1.8 eV which is in reasonable agreement with earlier measurement for Co$_3$O$_4$.[8] The $U$ term on Yb 4f electrons also pushed the filled states to lower energy range in the valence band enhancing the hybridisation with O 2p states. This arrangement for Yb 4f states agrees well with earlier DFT calculation for Gd$_{2-x}$Yb$_x$Zr$_2$O$_7$.[9] Figure S1(c) shows the PDOS calculated with slightly larger $U$ values; 3.3 eV for Co 3d electrons and 7.0 eV for Yb 4f electrons. Here, the $U$ value for Co equals to the recommended value from Materials Project[10] for Na$_1$CoO$_2$ and projector augmented-wave method.[11] calculations. These latter values did not alter the description of Co 3d state as both charge disproportionation, and the gap of ~1.8 eV between Co$^{3+}$ filled and empty states remained largely the same. We may, therefore, conclude that Materials Studio's default $U$ values suffice for an accurate physical description of the compound.

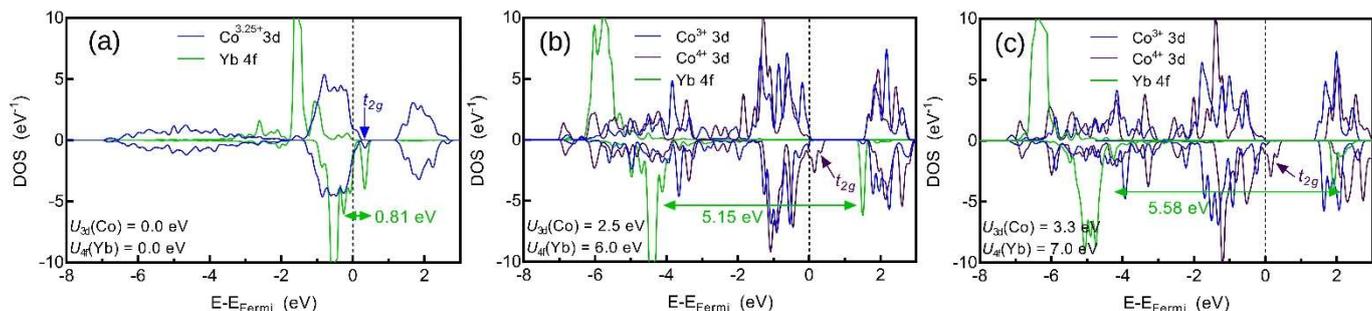

Figure S1. Partial density of state of $Na_{0.75}CoO_2:Yb_{Co}$ calculated with (a) LDA, (b) LDA+$U$ with $U_d(Co)$ = 2.5 eV, $U_f(Yb)$ = 6.0 eV and (c) with $U_d(Co)$ = 3.3 eV, $U_f(Yb)$ = 7.0 eV.

## The Effect of the Supercell Size

The calculations presented in the Article were conducted with supercells of $Na_{12}Co_{16}O_{32}$ supercell representing $Na_{0.75}CoO_2$ and $Na_8Co_{16}O_{32}$ supercell representing $Na_{0.50}CoO_2$ resulting in dopant to Co ratio of ~6.25%. We repeated the $E^f$ calculations of Gd doped $Na_{0.75}CoO_2$ using a larger supercell of $Na_{18}Co_{24}O_{48}$ composition, depicted in Figure S2, resulting in dopant to Co ratio of ~4.17%. As seen in Table S1, the $E^f$ values for the larger supercell are ~5% smaller indicating that achieving lower concentration doping is experimentally easier. More importantly, the sequence of stabilisation for Gd doping configurations was not altered by reducing doping concentration, implying that the conclusion drawn in the Article can be safely generalised for smaller doping concentrations.

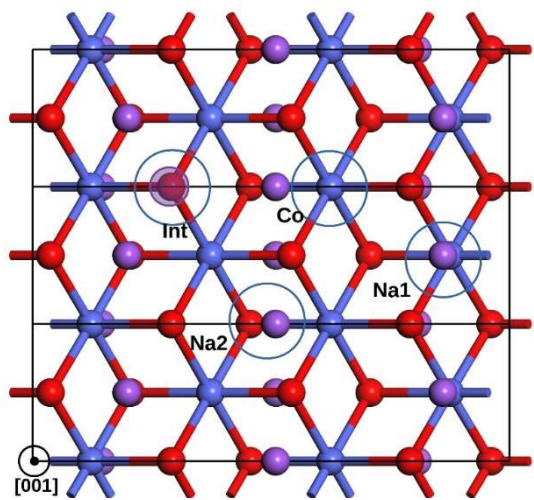

Figure S2. A top preview of the $Na_{0.75}CoO_2$ large supercell of $Na_{18}Co_{24}O_{48}$ composition.

Table S1. The formation energy of Gd dopant in $Na_{0.75}CoO_2$ in O rich and poor environments calculated with a larger supercell shown in Figure S2 and with the small supercell presented in the Article. The most stable configuration for each system is given in bold typeface.

|  | O Poor ($\mathcal{A}$) | | O Rich ($\mathcal{B}$) | |
| --- | --- | --- | --- | --- |
| Configuration | Large Supercell | Small Supercell | Large Supercell | Small Supercell |
| $Na_{0.75}CoO_2:Gd_{Co}$ | 5.31 | 5.51 | **4.75** | **4.95** |
| $Na_{0.75}CoO_2:Gd_{Int}$ | **4.94** | **5.13** | 10.70 | 10.91 |
| $Na_{0.75}CoO_2:Gd_{Na1}$ | 6.24 | 5.93 | 10.09 | 9.79 |
| $Na_{0.75}CoO_2:Gd_{Na2}$ | 6.38 | 6.02 | 10.24 | 9.88 |

## Spin Populations; Mulliken vs. Hirshfeld Schemes

The simulation presented in this work was performed with a plane wave (PW) basis set that, due to its delocalised nature, provides no information regarding the localisation of the electrons around the atomic centres. To obtain charge localisations, therefore, the charge density obtained with PW basis set needs to be projected onto a localised basis set. There are several ways to accomplish such a projection. One of the most popular schemes is Mulliken population analysis[12] which projects the charge densities onto pseudo-atomic orbitals, generated from the pseudopotentials used in the electronic structure calculation. These pseudo-atomic orbitals are calculated by solving for the lowest energy eigenstates of the pseudopotential in a sphere, using a spherical Bessel basis set. Mulliken population analysis, nonetheless, suffers from a high degree of sensitivity to the atomic basis set with which the charge densities were initially calculated,[13] rendering the analysis inconsistent at times. To examine the consistency of the Mulliken population analysis presented in the Article, a comprehensive comparison with Hirshfeld charge analysis was performed and presented in Table S2. Hirshfeld charges are defined relative to the deformation density, that is density difference between the molecular and unrelaxed atomic charge densities. Hirshfeld atomic populations provide a clear partitioning of the electron density which is less sensitive to the basis set than the Mulliken scheme.[14] According to Table S2, the difference between Mulliken and Hirshfeld spin populations (spin-up charge population minus spin-down charge population) is only few per cent at most, implying the validity and the consistency of the Mulliken population analysis presented in the Article.

Table S2. Charge population analysis for most stable doping configuration at O rich ($Na_{0.75}CoO_2$:$Gd_{Co}$, $Na_{0.75}CoO_2$:$Yb_{Co}$, $Na_{0.50}CoO_2$:$Gd_{Co}$, $Na_{0.50}CoO_2$:$Yb_{Co}$,) and O poor environments ($Na_{0.75}CoO_2$:$Gd_{Int}$, $Na_{0.75}CoO_2$:$Yb_{Na1}$, $Na_{0.50}CoO_2$:$Gd_{Int}$, $Na_{0.50}CoO_2$:$Yb_{Na2}$). Co's magnetisation refers to the sum of the magnitude of the spin density borne on all Co ions in the supercell and indicates the number of spin bearing $Co^{4+}$ ions.

| | Dopant's Magnetisation ($\hbar/2$) Mulliken | Dopant's Magnetisation ($\hbar/2$) Hirshfeld | Co's Magnetisation ($\hbar/2$) Mulliken | Co's Magnetisation ($\hbar/2$) Hirshfeld |
|---|---|---|---|---|
| $Na_{0.75}CoO_2$:$Gd_{Co}$ | 6.92 | 6.86 | 3.80 | 3.99 |
| $Na_{0.75}CoO_2$:$Gd_{Int}$ | 6.95 | 6.92 | 0.76 | 0.77 |
| $Na_{0.50}CoO_2$:$Gd_{Co}$ | 7.18 | 7.69 | 7.20 | 7.24 |
| $Na_{0.50}CoO_2$:$Gd_{Int}$ | 7.06 | 7.34 | 4.37 | 4.92 |
| $Na_{0.75}CoO_2$:$Yb_{Co}$ | 0.95 | 0.99 | 3.09 | 3.05 |
| $Na_{0.75}CoO_2$:$Yb_{Na1}$ | 0.50 | 0.51 | 1.96 | 1.92 |
| $Na_{0.50}CoO_2$:$Yb_{Co}$ | 0.95 | 1.00 | 6.12 | 6.13 |
| $Na_{0.50}CoO_2$:$Yb_{Na2}$ | 0.54 | 0.55 | 5.12 | 5.08 |

## Full Range of Chemical Potentials

In the Article, we presented the dopants' formation energy for two extreme points of the permissible $\Delta\mu$s; $\mathcal{A}$ representing O poor environment and $\mathcal{B}$ representing O rich environment. Here, present Yb's formation energy, presented in Table S2, for two additional points $\mathcal{A}'$ and $\mathcal{B}'$, marked in Figure S3, which still represent O poor and O rich environment but at the lowest possible $\Delta\mu_{Na}$ values. As seen in Table 2, $E^f$ values vary at most by ~2.5% along $\mathcal{A}\mathcal{A}'$ and $\mathcal{B}\mathcal{B}'$ edges. More importantly, the sequence of the stabilisation of Yb doping configuration remains the same. The case is the same for Gd dopant as well.

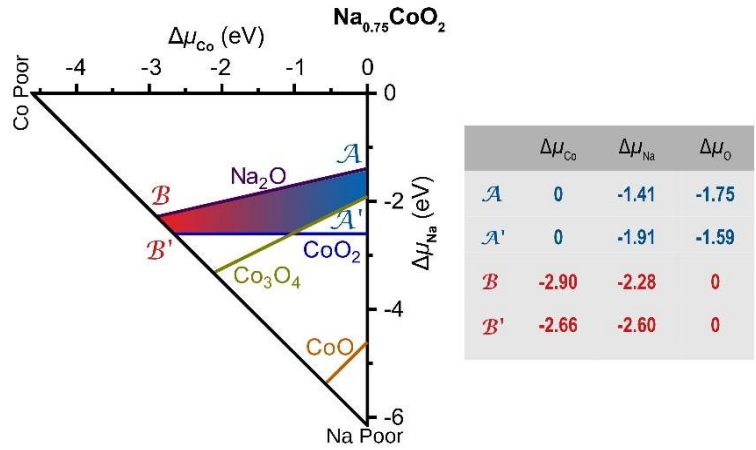

Figure S3. The accessible ($\Delta\mu$) chemical potential range. The triangle vertices are determined by the formation enthalpy of $Na_{0.75}CoO_2$. Limits are imposed by the formation of competing $Na_2O$, $Co_3O_4$, $CoO_2$. $\Delta\mu$ values are given for points $\mathcal{A}$, $\mathcal{A}'$, $\mathcal{B}$ and $\mathcal{B}'$ in eV.

Table S3. The formation energy in eV of Yb dopant in $Na_{0.75}CoO_2$. Notions $\mathcal{A}$, $\mathcal{A}'$, $\mathcal{B}$ and $\mathcal{B}'$ correspond to chemical potential extremes presented in Figure S3.

| Configuration | $\mathcal{A}$ | $\mathcal{A}'$ | $\mathcal{B}$ | $\mathcal{B}'$ |
|---|---|---|---|---|
| $Na_{0.75}CoO_2$:$Yb_{Co}$ | 14.47 | 14.63 | **13.32** | **13.55** |
| $Na_{0.75}CoO_2$:$Yb_{Int}$ | 14.69 | 14.85 | 16.44 | 16.44 |
| $Na_{0.75}CoO_2$:$Yb_{Na1}$ | **14.03** | **13.69** | 14.91 | 14.60 |
| $Na_{0.75}CoO_2$:$Yb_{Na2}$ | 14.42 | 14.08 | 15.30 | 14.99 |

## The Origin of Koshibae's Equation

In an attempt to reach a new equilibrium under temperature gradient, the flow of electrons from the hot side to the cold side in a solid inevitably causes an electric current in metals and semiconductors. These electrons would also carry heat and therefore, entropy when experiencing a temperature gradient that tips off the thermal equilibrium. The Seebeck coefficient which is defined as the ratio of the generated potential difference between the hot and cold sides per a given temperature difference can be generally expressed as:

$$S = \frac{k_B}{e}\left(\frac{\partial s}{\partial n}\right)_{T,V} = -\frac{k_B}{e}\left(\frac{\partial \mu}{\partial T}\right)_{V,n}, \quad \text{Equation S1}$$

where $s$ is the entropy density, $n$ the density of the number of electrons, $V$ the volume, $\mu$ the chemical potential of electrons, $k_B$ Boltzmann's constant and where the second equation follows from a Maxwell relation.[15] In the high-temperature regime, $S$ can be expressed with a temperature-independent Heikes formula:[16, 17]

$$S = -\frac{k_B}{e} \text{Ln}\left(2\frac{1-\rho}{\rho}\right). \qquad \text{Equation S2}$$

In which $\rho$ is charge carrier density. The factor 2, in Equation S2, signifies the electronic degeneracy parameter of a single electron. In the case where electron hopping between hetero-valent transition metal ions, say ions A and B, is the sole conduction mechanism, Heike's formula has to be revised into Koshibae's equation to accommodate the additional entropy factors imposed by the orbital and spin degeneracies of the d electrons:[18, 19]

$$S(T \to \infty) = -\frac{k_B}{e} \ln\left[\frac{g(A)}{g(B)} \frac{n_A}{n_B}\right]. \qquad \text{Equation S3}$$

Here, $g$ equals to the different possible ways in which electrons can be arranged in the orbitals of A and B ions. In other words, $g$ refers to the degeneracy of the electronic configuration in ions A and B. Finally, $n$ is the concentration of a given ion. This concept is illustrated in Figure S4. In $Na_xCoO_2$, the six electrons of the low-spin $Co^{3+}$ constitute a $t_{2g}^6 e_g^0$ configuration for which there is no other degenerate state, that is $S$(entropy) = $\text{Ln}\{g(Co^{3+})\}$ = Ln1 = 0. In contrast, the low-spin $Co^{4+}$ ion, since one electron is removed, there are six degenerate configurations. Consequently, $S$(entropy) = $\text{Ln}\{g(Co^{4+})\}$ = Ln6. As an electron hops from a $Co^{3+}$ ion to a $Co^{4+}$ ion, an entropy flux moves in the opposite direction of the electric current; this phenomenon is referred to as spin entropy flow.[20]

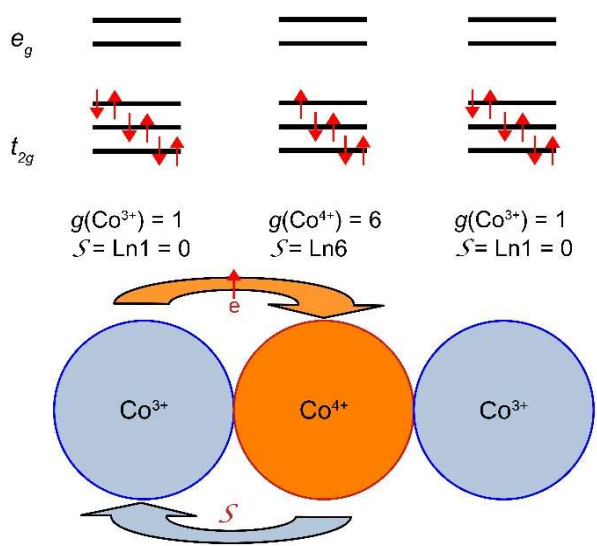

Figure S4. The schematics of spin entropy flow in $Na_xCoO_2$. $S$ and $g$ are the entropy and the electronic degeneracy per site, respectively.

## Calculation of Transport Properties

The carrier concentration ($n$) was calculated by dividing the number of $Co^{4+}$ ions which each bears a single hole by the cell volume 622.031172 Å$^3$. The conductivity ($\sigma$) then was calculated as $\sigma = e \cdot n \cdot m$ in which $e$ and $m$ are electron's charge and carrier mobility, respectively. $m$ was approximated as 1.0 cm$^2$V$^{-1}$s$^{-1}$ for all compounds.

VOR: `https://doi.org/10.1088/1361-648X/ab5bdb`

Table S4. Tabulated transport properties calculated for $Na_{0.75}CoO_2$ and $Na_{0.50}CoO_2$ doped with $M^{3+}$ rare-earth dopant (Gd/Yb) at a concentration of M/Co = ~6.25%. Available experimental data are provided in the Table's footnotes for comparisons.

| System | Composition | Seebeck Coefficient (μV/K) | holes per supercell | Carrier concentration (cm$^{-3}$) | Conductivity ($\Omega^{-1}$cm$^{-1}$) | Power factor (mWm$^{-1}$K$^{-2}$) |
|---|---|---|---|---|---|---|
| $Na_{0.75}CoO_2$ | $Na_{12}Co_{16}O_{32}$ | 249[a] | 4 | 6.43×10$^{+21}$ | 1.03×10$^{+3}$ | 6.392[b] |
| $Na_{0.75}CoO_2$:$M_{Co}$ | $Na_{12}Co_{15}MO_{32}$ | 242 | 4 | 6.43×10$^{+21}$ | 1.03×10$^{+3}$ | 6.013 |
| $Na_{0.75}CoO_2$:$M_{Na}$ | $Na_{11}Co_{16}MO_{32}$ | 322 | 2 | 3.22×10$^{+21}$ | 5.15×10$^{+2}$ | 5.344 |
| $Na_{0.75}CoO_2$:$M_{Int}$ | $Na_{12}Co_{16}MO_{32}$ | 388 | 1 | 1.61×10$^{+21}$ | 2.58×10$^{+2}$ | 3.873 |
| $Na_{0.50}CoO_2$ | $Na_8Co_{16}O_{32}$ | 154[c] | 8 | 1.29×10$^{+22}$ | 2.06×10$^{+3}$ | 4.912[d] |
| $Na_{0.50}CoO_2$:$M_{Co}$ | $Na_8Co_{15}MO_{32}$ | 143 | 8 | 1.29×10$^{+22}$ | 2.06×10$^{+3}$ | 4.208 |
| $Na_{0.50}CoO_2$:$M_{Na}$ | $Na_7Co_{16}MO_{32}$ | 198 | 7 | 9.65×10$^{+21}$ | 1.55×10$^{+3}$ | 6.085 |
| $Na_{0.50}CoO_2$:$M_{Int}$ | $Na_8Co_{16}MO_{32}$ | 222 | 5 | 8.04×10$^{+21}$ | 1.29×10$^{+3}$ | 6.367 |

[a] ~200 μV/K at 800 K, Ref. 21.
[b] ~7.69 mWm$^{-1}$K$^{-2}$ at 800 K, Ref. 21.
[c] ~130 μV/K at 800 K, Ref. 22.
[d] ~5.0 mWm$^{-1}$K$^{-2}$ at 300 K, Ref. 23.